\documentclass[aps,prl,twocolumn,superscriptaddress,amsmath,floatfix,showpacs]{revtex4} 
\usepackage{graphics,graphicx,epsfig,color}

\def\G0W0{$\mathrm G^\circ\mathrm W^\circ$}
\def\barPhi{\overline{\Phi}}
\begin{document}
\title{
  Bridging the size gap between density-functional and many-body
  perturbation theory
}

\author{P. Umari}
\affiliation{
  INFM-CNR DEMOCRITOS Theory@Elettra group, c/o Sincrotrone Trieste,
  Area Science Park, I-34012 Basovizza, Trieste, Italy 
} 

\author{Geoffrey Stenuit}
\affiliation{
  INFM-CNR DEMOCRITOS Theory@Elettra group, c/o Sincrotrone Trieste,
  Area Science Park, I-34012 Basovizza, Trieste, Italy 
} 

\author{Stefano Baroni}
\affiliation{
  SISSA -- Scuola Internazionale Superiore di Studi Avanzati, via
  Beirut 2-4, I-34014 Trieste Grignano, Italy
}
\affiliation{
  INFM-CNR DEMOCRITOS Theory@Elettra group, c/o Sincrotrone Trieste,
  Area Science Park, I-34012 Basovizza, Trieste, Italy 
} 

\date{\today}

\begin{abstract}
  The calculation of quasi-particle spectra based on the GW
  approximation is extended to systems of hundreds of atoms, thus
  expanding the size range of current approaches by more than one
  order of magnitude. This is achieved through an optimal strategy,
  based on the use of Wannier-like orbitals, for evaluating the
  polarization propagator. Our method is validated by
  calculating the vertical ionization energies of the benzene molecule
  and the band structure of crystalline silicon. Its potentials are
  then demonstrated by addressing the quasi-particle spectrum of a
  model structure of vitreous silica, as well as of the
  tetraphenylporphyrin molecule.
\end{abstract}


\maketitle

Density-functional theory (DFT) has grown into a powerful tool for the
numerical simulation of matter at the nanoscale, allowing one to study
the structure and dynamics of realistic models of materials consisting
of up to a few thousands atoms, these days \cite{dft}. The scope of
standard DFT, however, is limited to those dynamical processes that do
not involve electronic excitations. The most elementary such
excitation is the removal/addition of an electron from a system
originally in its ground state. These processes are accessible to
direct/inverse photo-emission spectroscopies and can be described in
terms of {\em quasi-particle} (QP) spectra \cite{hl69}. In insulators,
the energy difference between the lowest-lying quasi-electron state
and the highest-lying quasi-hole state is the QP band gap, a quantity
that is severely (and to some extent erratically) underestimated by
DFT \cite{ag98}.

Many-body perturbation theory (MBPT), in turn, provides a general,
though unwieldy, framework for QP and other excitation (such as
optical) spectra \cite{hl69}. A numerically viable approach to QP
energy levels (known as the GW approximation, GWA) was introduced in
the 60's \cite{h65}, but it took two decades for a realistic
application of it to appear \cite{hl85}, and even today the numerical
effort required by MBPT is such that its scope is limited to systems
of a few handfuls of atoms.  Even so, and in spite of the success met
by MBPT in real materials \cite{orr01}, the approximations made for
the most demanding of its applications are such as to shed some
legitimate doubts on their general applicability. This situation will
be referred to as the {\em size gap} of MBPT calculations.

In this letter we present a strategy to substantially reduce the size
gap of MBPT, based on the adoption of Wannier-like orbitals
\cite{mv97,smv01,gfs03} to represent the response functions whose
calculation is the main size-limiting factor of MBPT. Although we
focus on QP spectra within the GWA, this strategy easily generalizes
to optical spectra, as calculated from the Bethe-Salpeter equation
\cite{bse}. Our method is benchmarked by the calculation of the
ionization potentials of the benzene molecule and of the band
structure of crystalline silicon, and its potentials demonstrated by
case calculations on vitreous silica and on the free-base
tetraphenylporphyrin molecule (TPPH$_2$).

QP energies (QPE) are eigenvalues of a Schr\" odinger-like equation
(QPEq) for the so-called QP amplitudes (QPA), which is similar to the
DFT Kohn-Sham equation with the exchange-correlation potential,
$V_{xc}(\mathbf{r})$, replaced by the non-local, energy-dependent, and
non-Hermitian self-energy operator, $\tilde\Sigma(\mathbf{r},
\mathbf{r}',E)$ (a tilde indicates the Fourier transform of a
time-dependent function). Setting $\tilde\Sigma(\mathbf{r},
\mathbf{r}';E) = - {\rho(\mathbf{r},\mathbf{r}') \over |
  \mathbf{r}-\mathbf{r}' | }$ ($\rho$ being the one-particle density
matrix) would turn the QPEq into the Hartree-Fock equation. The next
level of approximation is the GWA \cite{h65} where $\tilde\Sigma$ is
the product of the one-electron propagator, $G$,\ and of the
dynamically screened interaction, $W$:
\begin{equation}
  \label{eq:sigma-GW}
  \Sigma_{GW}(\mathbf{r},\mathbf{r}';t-t') = i
  G(\mathbf{r},\mathbf{r}';t-t') 
   W(\mathbf{r},\mathbf{r}';t-t'),
\end{equation}
where $ W = v+ v\cdot\Pi \cdot v$, $\Pi (\mathbf{r},\mathbf{r}';t-t')
= {\delta n(\mathbf{r},t) \over \delta V(\mathbf{r}',t')} = (1-P\cdot
v)^{-1}\cdot P$ is the reducible electron polarization propagator
(polarizability), $P$ its irreducible counterpart,
$v(\mathbf{r},\mathbf{r}';t-t') = {1 \over
  |\mathbf{r}-\mathbf{r}'|}\delta(t-t')$ is the bare Coulomb
interaction, $n$ and $V$ are the electron density distribution and
external potential, respectively, and a dot indicates the product of
two operators, such as in $v \cdot \chi(\mathbf{r},\mathbf{r}',t-t') =
\int d\mathbf{r}'' dt'' v(\mathbf{r},\mathbf{r}'';t-t'')
\chi(\mathbf{r}'',\mathbf{r}';t''-t')$.

The GWA alone does not permit to solve the QPEq, unless $G$ and $W$
are known, possibly depending on the solution of the QPEq itself. One
of the most popular further approximations is the so called \G0W0
approximation (\G0W0{A}), where the one-electron propagator is
obtained from the QPEq using a model real and energy-independent
self-energy, such as {\em e.g.}  $\tilde\Sigma^\circ=V_{\rm
  xc}(\mathbf{r})\delta(\mathbf{r}-\mathbf{r}')$, and the irreducible
polarizability is calculated in the random-phase approximation (RPA):
$G^\circ(\mathbf{r},\mathbf{r}';\tau) = i \sum_v \psi_v(\mathbf{r})
\psi_v^*(\mathbf{r}') \mathrm{e}^{-i\epsilon_v \tau}\theta(-\tau) -i
\sum_c \psi_c(\mathbf{r}) \psi_c^*(\mathbf{r}')
\mathrm{e}^{-i\epsilon_c \tau} \theta(\tau) $ ($\psi$ and $\epsilon$
are zero-th order QPAs and QPEs, referred to the Fermi energy, $v$ and
$c$ suffixes indicate states below and above the Fermi energy,
respectively, and $\theta$ is the Heaviside step function) and
$P^\circ(\mathbf{r},\mathbf{r}';\tau) = -iG^\circ(\mathbf{r},\mathbf{r}';
\tau) G^\circ(\mathbf{r}',\mathbf{r};-\tau) $. To first order in
$\Sigma'=\Sigma_{G^\circ W^\circ}-\Sigma^\circ$, QPEs are given by the
equation:
\begin{equation}
  \label{eq:QPE}
  E_n\approx \epsilon_n+\langle \tilde\Sigma_{G^\circ W^\circ}(E_n) 
  \rangle_n - \langle V_{XC} \rangle_n,
\end{equation}
where $\langle A\rangle_n=\langle\psi_n| A|\psi_n \rangle $.

The apparently simple \G0W0A still involves severe difficulties,
mainly related to the calculation and manipulation of the
polarizability that enters the definition of $W^\circ$. These
difficulties are often addressed using the so called plasmon-pole
approximation \cite{hl85}, which however introduces noticeable
ambiguities and inaccuracies when applied to inhomogeneous systems
\cite{srg08}. A well established technique to address QP spectra in
real materials without any crude approximations on response functions
is the {\em space-time method} (STM) by Godby {\em et al.}
\cite{rgn95}. In the STM the time/energy dependence of the \G0W0
operators is represented on the imaginary axis, thus making them
smooth (in the frequency domain) or exponentially decaying (in the
time domain).  The various operators are represented on a real-space
grid, a choice which is straightforward, but impractical for systems
larger than a few handfuls of inequivalent atoms. In this paper we
combine the imaginary time/frequency approach of the STM with a novel
representation of the response functions, based on localized
Wannier-like orbitals, thus enhancing the scope of MBPT calculations
so as to embrace systems potentially as large as a few hundreds atoms.

In the STM, the self-energy expectation value in Eq. \eqref{eq:QPE} is
obtained by analytically continuing to the real frequency axis the
Fourier transform of the expression:
\begin{multline}
  \label{eq:Sigman}
  \langle\Sigma_{G^\circ W^\circ}(i\tau) \rangle_n = 
  \mp \sum_l \mathrm{e}^{\epsilon_l\tau} \times \\ \int
  \psi_n(\mathbf{r})\psi_l(\mathbf{r})\psi_l(\mathbf{r}')
  \psi_n(\mathbf{r}') W(\mathbf{r},\mathbf{r}';i\tau) 
  d\mathbf{r} d\mathbf{r}', 
\end{multline}
where the upper (lower) sign holds for positive (negative) times, the
sum extends below (above) the Fermi energy, and QPAs are assumed to be
real. By substituting $v$ for $W$, Eq. \eqref{eq:Sigman} yields the
exchange self-energy, whereas $v\cdot \Pi\cdot v$ yields the
correlation contribution, $\Sigma_C$, whose evaluation is the main
size-limiting step of GW calculations.

Suppose that a small, time-independent, basis set to represent the
polarizability exists: $\Pi(\mathbf{r},\mathbf{r}',i\tau) \approx
\sum_{\mu\nu} \Pi_{\mu\nu}(i\tau) \barPhi_\mu(\mathbf{r})
\barPhi_\nu(\mathbf{r}')$.  Eq. \eqref{eq:Sigman} then gives:
\begin{equation}
  \label{eq:SigmaCn}
  \langle\Sigma_C(i\tau)\rangle_n \approx \mp \sum_{l\mu\nu}
  \mathrm{e}^{\epsilon_l\tau} \Pi_{\mu\nu}(i\tau)
  S_{nl,\mu}S_{nl,\nu} \theta(E^1_C-\epsilon_l),
\end{equation}
where $S_{nl,\nu}=\int \psi_n(\mathbf{r})\psi_l(\mathbf{r}) \frac{1}{|
  \mathbf{r} - \mathbf{r}'|}
\barPhi_\nu(\mathbf{r}')d\mathbf{r}d\mathbf{r}'$ and $E_C^1$ is an
energy cutoff that limits the number of conduction states to be used
in the calculation of the self-energy.  A convenient representation of
the polarizibility would thus allow QPEs to be calculated from
Eq. \eqref{eq:QPE}, by analytically continuing to the real axis the
Fourier transform of Eq. \eqref{eq:SigmaCn}. Such an optimal
representation is identified in three steps: {\em i)} we first express
the Kohn-Sham orbitals, whose products enter the definition of
$P^\circ$, in terms of localized, Wannier-like, orbitals; {\em ii)} we
then construct a basis set of localized functions for the manifold
spanned by products of Wannier orbitals; finally, {\em iii)} this
basis is further restricted to the set of eigenvectors of $P^\circ$,
corresponding to eigenvalues larger than a given threshold.

Let us start from the RPA irreducible polarizability:
\begin{equation}  \label{eq:Prpa}
  \tilde P^\circ(\mathbf{r},\mathbf{r}';i\omega) = \sum_{cv}
  \Phi_{cv}(\mathbf{r}) \Phi_{cv}(\mathbf{r}')
  \tilde\chi^\circ_{cv}(i\omega), 
\end{equation}
where $ \tilde\chi^\circ_{cv}(i\omega)=2\mathrm{Re}\left ( {1\over
    i\omega-\epsilon_c+\epsilon_v} \right )$ and $
\Phi_{cv}(\mathbf{r}) = \psi_c(\mathbf{r})\psi_v(\mathbf{r})$.  We
express valence and conduction QPAs in terms of localized,
Wannier-like, orbitals:
\begin{equation}
  \label{eq:W-rotate}
  \begin{split}
    u_{s}(\mathbf{r}) &= \sum_{v} \mathcal{U}^{-1}_{sv}
    \psi_{v}(\mathbf{r}) \theta(-\epsilon_{v}) \\ 
    v_{s}(\mathbf{r}) &= \sum_{c} \mathcal{V}^{-1}_{sc}
    \psi_{c}(\mathbf{r}) \theta(\epsilon_{c})
    \theta(E^2_C-\epsilon_{c}), 
  \end{split}
\end{equation}
where $E^2_C \le E^1_C$ is a second energy cutoff that limits a {\em lower
  conduction manifold} (LCM) to be used in the construction of the
polarization basis. According to the choice of the $\mathcal{U}$ and
$\mathcal{V}$ matrices, the $u$'s and $v$'s can be either maximally
localized \cite{mv97,gfs03} or non-orthogonal generalized \cite{rw05}
Wannier functions. We then reduce the number of product functions from
the product between the number of valence and conduction states, which
scales quadratically with the system size, to a number that scales
linearly. To this end, we express the $\Phi$'s as approximate linear
combinations of products of the $u$'s $v$'s: $ \Phi_{cv}(\mathbf{r})
\approx \sum_{rs} \mathcal{O}_{cv,rs} W_{rs}(\mathbf{r})
\theta(|W_{rs}|^2-s_1)$, where $ \mathcal{O}_{cv,c'v'}=
\mathcal{U}_{vv'} \mathcal{V}_{cc'} $, $ W_{rs}(\mathbf{r})=
u_{r}(\mathbf{r}) v_{s}(\mathbf{r})$, $|W_{cv}|$ is the $L^2$ norm of
$W_{rs}(\mathbf{r})$, which is arbitrarily small when the centers of
the $u_r$ and $v_s$ functions are sufficiently distant, and $s_1$ is
an appropriate cutoff. The number of products can be further reduced
on account of the non-orthogonality and mutual linear dependence of
the $W$'s. To this end, let us define the overlap matrix: $
\mathcal{Q}_{\rho\sigma}=\int W_\rho(\mathbf{r}) W_\sigma(\mathbf{r})
d\mathbf{r}$, where the $\rho$ and $\sigma$ indices stand for pairs of
$rs$ indices.  The magnitude of the eigenvalues is a measure of linear
dependence, and an orthonormal basis can be obtained by retaining only
those eigenvectors $\mathcal{U}_\nu$ whose eigenvalue, $q_{\nu}$, is
larger than a given threshold, $s_2$: $\overline{\Phi}_\nu(\mathbf{r})
\approx \frac{1}{\sqrt{q_\nu}} \sum_{\rho} \mathcal{U}_{\nu\rho}
W_{\rho}(\mathbf{r})$, for $q_\nu > s_2$. A final (and practically
ultimate) refinement can be achieved by diagonalizing $P^\circ$ in the
basis thus reduced and by retaining only those eigenvectors whose
eigenvalue is larger that a third threshold, $s_3$.  The result of
this last procedure would depend on frequency, which would make it
impractical. We have verified, however, that the manifold spanned by
the most important eigenvectors of $P^\circ$ {\em in the (imaginary)
  time domain} depends very little on time, which permits the use of a
same basis at different frequencies. When the
basis for $P^\circ$ is built out of {\em orthonormal} Wannier
orbitals, this last refinement does not result in any further
improvement if $s_3\ge s_2$ \cite{XQ}.
  
Once an optimal basis set has been thus identified, an
explicit representation for the irreducible polarizability,
\begin{equation} 
  \label{eq:Prepr} \tilde P^\circ(\mathbf{r},\mathbf{r},i\omega) =
  \sum_{\mu\nu} \tilde P^\circ_{\mu\nu}(i\omega)
  \overline{\Phi}_\mu(\mathbf{r})   \overline{\Phi}_\nu(\mathbf{r}'), 
\end{equation} 
is obtained. By equating Eq. \eqref{eq:Prpa} to Eq. \eqref{eq:Prepr}
and taking into account the orthonormality of the $\overline\Phi$'s,
one obtains:
\begin{equation}
\label{eq:PolaCn}
  P^\circ_{\nu\mu}(i\omega)=\sum_{cv} T_{cv,\mu}T_{cv,\nu}\tilde
  \chi^\circ_{cv}(i\omega) \theta(E^1_C-\epsilon_c),
\end{equation}
where $T_{cv,\mu}=\int \Phi_{cv}(\mathbf{r})
\overline{\Phi}_\mu(\mathbf{r}) 
d\mathbf{r}$.
A representation for $\Pi$ is finally obtained by simple matrix
manipulations.

\begin{figure}
  \includegraphics[angle=-90.,width=8.0cm]{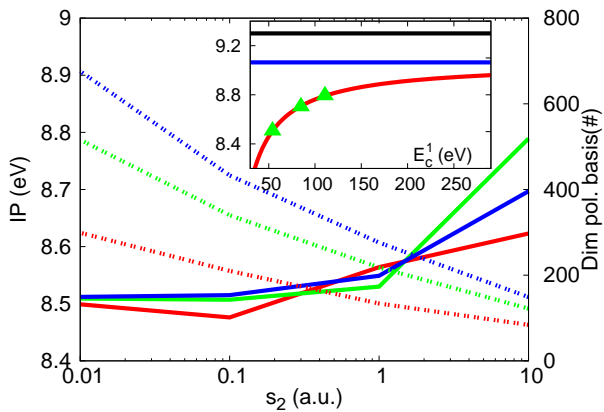}
  \caption{ 
    \label{fig1}
    Calculated ionization potential of the benzene molecule (solid
    lines, left scale) and dimension of the polarization basis (dashed
    lines, right scale) versus the $s_2$ cutoff. The polarization
    basis has been constructed with a conduction energy cutoff
    $E^2_C=16.7~\mathrm{eV}$ (red, 100 states), 
    $E^2_C=28.6~\mathrm{eV}$ (green, 300 states), and
    $E^2_C=38.3~\mathrm{eV}$ (blue, 500 states). Inset: calculated
    ionization potential as a function of the overall conduction
    energy cutoff, $E^1_C$. Black line: experimental value; red
    line: fit to the calculated values (green triangles);
    blue line extrapolated value. See text for more details. 
  }
\end{figure}

Our scheme has been implemented in the {\sc Quantum ESPRESSO} density
functional package \cite{qe}, for norm-conserving as well as
ultra-soft pseudopotentials, resulting in a new module called {\em
  gww.x} which uses a Gauss-Legendre discretization of the imaginary
time/frequencies half-axes, and that is parallelized accordingly.
We first illustrate our scheme by considering an isolated benzene molecule
% \cite{benzpar}
in a periodically repeated cubic cell with an edge of $20$ a.u. using
a first conduction energy cutoff $E^1_C=56.7~\mathrm{eV}$,
corresponding to 1000 conduction states, and a cutoff on the norm of
Wannier products $s_1=0.1~\mathrm{a.u.}$ \cite{technicalities}.  In
Fig.~1 we display the dependence of the calculated ionization
potential (IP) on the second conduction energy cutoff used to define
the polarization basis, $E^2_C$, and on the cutoff on the eigenvalues
of the overlap matrix between Wannier products, $s_2$.  Convergence
within $0.01$ eV is achieved with a conduction energy cutoff $E_C^2$
smaller than 30 eV (less than 300 states) and a polarization basis set
of only $\sim 400$ elements. The convergence of other QPEs is
similar. The inset of Fig. \ref{fig1}, shows the convergence of the IP
with respect to $E^1_C$, which turns out to be unexpectedly
slow. These data can be accurately fitted by the simple formula
$\mathrm{IP}(E^1_C)=\mathrm{IP}(\infty)+A/E^1_C$, resulting in a
predicted ionization potential $\mathrm{IP (\infty)= 9.1~eV}$, in good
agreement with the experimental value of $9.3$ eV \cite{ldp76}. The
potential of our method for large molecular system is illustrated by a
calculation for the TPPH$_2$ molecule ($\mathrm{C_{44}H_{30}N_4}$) in
a periodically repeated orthorhombic supercell of $75.6\times 75.6
\times 26.5$ a.\ u.\ \cite{technicalities}.  Using values of $31.1$,
$40.5$ and $48.1$ eV for $E_C^1$ (corresponding to $2000$, $3000$ and
$4000$ conduction states) and $E_C^2=17.2~\mathrm{eV}$ (corresponding
to $750$ conduction states), $s_1=1.5$ and $s_2=0.01~\mathrm{a.u.}$,
which lead to a basis of $2797$ elements, yields an extrapolated
$\mathrm{IP} (\infty)$ of $6.0$ eV, in fair agreement with the
experimental value of $6.4$ eV \cite{glo99}.

\begin{table}
  \begin{tabular}{lcccc}
    \hline
    &  Th$_1$    &   Th$_2$    &  prev th & Expt\ \\
    \hline
    $N_P$         &  4847  &  6510   &        & \\                     
    \hline
    $\Gamma_{\mathrm 1v}$   & -11.45 & -11.49  &  -11.57 & -12.5$\pm$0.6\\
    X$_{\mathrm 1v}$        & -7.56  & -7.58   &  -7.67  &  \\ 
    X$_{\mathrm 4v}$        & -2.79  & -2.80   &  -2.80  & -2.9, -3.3$\pm$0.2\\
    $\Gamma'_{\mathrm 25c}$ &  0.    &   0.    &   0.    & 0. \\
X$_{\mathrm 1c}$        &  1.39  &  1.41   &   1.34  &  1.25 \\
$\Gamma'_{\mathrm 15c}$ &  3.22  &  3.24   &   3.24  & 3.40,3.05\\
$\Gamma'_{\mathrm 2c}$  &  3.87  &  3.89   &   3.94  & 4.23, 4.1 \\
\hline
\end{tabular}
\caption{
  \label{tab1}
  QPEs (eV) calculated in crystalline silicon and compared with
  experimental (as quoted in Ref. \onlinecite{rgn95}) and previous
  theoretical results \cite{rgn95}. `Th$_1$' and `Th$_2$' indicate
  calculations made with $s_2=0.01$ and $s_2=0.001$ a.u. respectively,
  while $N_P$ is the dimension of the polarization basis.
}
\end{table}

\begin{figure}
  \includegraphics[width=7.0cm]{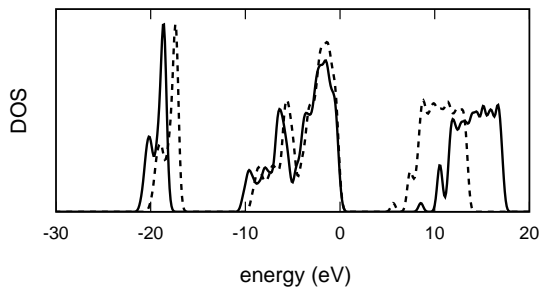}
  \caption{
    \label{fig2}
    Electronic density of states for a model of vitreous silica: LDA
    (dashed line) and GW (solid line). A Gaussian broadening of
    0.25 eV has been used.
  }
\end{figure}

In order to demonstrate our scheme for extended systems
\cite{extended}, we consider crystalline silicon treated using a
64-atom simple cubic cell \cite{technicalities} at the experimental
lattice constant and sampling the corresponding Brillouin zone (BZ)
using the $\Gamma$ point only. This gives the same sampling of the
electronic states as would result from $6$ points in the irreducible
wedge of the BZ of the elementary 2-atom unit cell.  Our calculations
were performed using $E_C^1=94.6~\mathrm{eV}$ (corresponding to 3200
conduction states) and $E_C^2=33.8~\mathrm{eV}$ (corresponding to 800
states in the LCM), $s_1=1.0~\mathrm{a.u.}$ and two distinct values
for $s_2$ (0.01 and 0.001). In Tab.\ \ref{tab1} we summarize our
results and compare them with previous theoretical results, as well as
with experiments.  An overall convergence within a few tens meV is
achieved with a $s_2$ cutoff of $0.001$ a.u., corresponding to a
polarization basis of $\sim 6500$ elements. The residual small
discrepancy with respect to previous results \cite{rgn95} is likely
due to our use of a supercell, rather
than the more accurate k-point sampling used in previous work.  Our
ability to treat large supercells give us the possibility to deal with
disordered systems that could hardly be addressed using conventional
approaches. In Fig.\ \ref{fig2} we show the QPE density of states as
calculated for a 72-atom model of vitreous silica
\cite{spc95,technicalities}. We used $E_C^1=48.8~\mathrm{eV}$
(corresponding to $1000$ conduction states), $E_C^2=30.2~\mathrm{eV}$
(corresponding to $500$ states in the LCM), $s_1=1$ a.u. and $s_2=0.1$
a.u. (giving rise to a polarization basis of $3152$ elements). We
checked the convergence with respect to the polarization basis by
considering $s_2=0.01$ a.u. which leads to a basis of $3933$
elements. Indeed, the calculated QPEs differ in average by only $0.01$,
eV with a maximum discrepancy of $0.07$ eV \cite{sio2beta}. The
quasi-particle band-gap resulting from our calculations is $8.5$ eV, to
be compared with an experimental value of $\sim 9$ eV \cite{sio2exp}
and with a significantly lower value predicted by DFT in the
local-density approximation ($5.6$ eV).

In conclusion, we believe that expressing density response functions
in terms of localized basis sets will permit to extend the scope of
many-body perturbation theory to large models of molecular and
extended, possibly disordered, systems. The extension of the
methodology presented in this paper for quasi-particle spectra to
optical spectroscopies using the Bethe-Salpeter formalism is
straightforward and presently under way.

We thank C. Cavazzoni for his help in the parallelization
of the code. This work has been partially funded under the
Italian CNR-INFM {\em Seed Projects} scheme.

\bibliographystyle{PhysRev}

\end{document}